\documentclass[aps,floats,twocolumn,showpacs]{revtex4}
\usepackage{amssymb}
\usepackage{amsmath}
\usepackage{graphicx}
 \def\bc{\begin{center}}          \def\ec{\end{center}}
 \allowdisplaybreaks

\begin{document}
\title{Plasma Wakefield Acceleration with a Modulated Proton Bunch}
\author{A. Caldwell}
\affiliation{Max-Planck-Institut f\"ur Physik, 80805, M\"unchen, Germany}
\author{K. V. Lotov}
\affiliation{Budker Institute of Nuclear Physics, 630090, Novosibirsk, Russia
\\
 Novosibirsk State University, 630090, Novosibirsk, Russia}
 \date{\today}
\begin{abstract}
The plasma wakefield amplitudes which could be achieved via the
modulation of a long proton bunch are investigated.  We find that in the limit of long bunches compared to the plasma
wavelength, the strength of the accelerating fields is directly proportional to
the number of particles in the drive bunch and inversely proportional to the square of the transverse bunch size. The scaling laws were tested and verified in detailed simulations using parameters of existing proton accelerators, and large electric fields were achieved, reaching $1$~GV/m for LHC bunches.  Energy gains for test electrons beyond $6$~TeV were found in this case.
\end{abstract}
\maketitle

Proton-driven plasma-wakefield acceleration has been proposed recently as a
means of accelerating bunches of electrons to high
energies~\cite{NatPhys9-363,PRST-AB13-041301}. The basic idea is to use a
plasma to transfer the energy from a bunch of protons, which can be accelerated
to the multi-TeV regime today in existing accelerators, to a bunch of
electrons, for which energies have so far been limited to about 100~GeV due to
synchrotron radiation losses in circular accelerators.  In
\cite{NatPhys9-363,PRST-AB13-041301}, the proton bunch was assumed to have been
compressed to an rms length of $100$~$\mu$m to achieve the strong accelerating
fields in the plasma.  Such short proton bunches are not available today, and
compression schemes, while conceivable~\cite{IPAC10-4392,AIP1229-510}, will
require long distances and substantial RF power, and will therefore be costly.
A possible alternative to bunch compression is to start with an existing long
proton bunch and divide it up into a series of microbunches in a plasma. A~long
proton bunch propagating in a plasma will naturally be modulated at the plasma
wavelength, and the modulated bunch will in turn set up oscillations of the
plasma electrons.  The modulation of the bunch grows rapidly from noise and the
growth rate has recently been studied theoretically~\cite{ref:KP}, with
calculations showing that the modulation can be very effectively produced.

We report here both parametric studies and simulation results for the electric
fields which can be achieved via this mechanism, and discuss achievable energy
gains using CERN PS, SPS and LHC bunch parameters as concrete examples.  We
start with a brief review of the relevant formulae for the axial electric
fields produced by bunches of charged particles in a plasma in the linear
regime.  This is followed by a discussion on the conditions necessary to reach
the regime of strong modulations, and simulation results are shown depicting
this growth of modulations.  We then discuss the conditions necessary for
reaching maximum accelerating fields, and show comparisons of our calculations
with simulations.  Finally, we discuss the limitations on energy gain for
bunches of electrons and summarize the results for different driving bunch
parameters.

\section{Single bunch}
For Gaussian shaped drive bunches in the linear regime (where the bunch charge
density is much lower than the plasma electron density), the maximum axial
electric field is given by~\cite{ref:WeiLu}
\begin{eqnarray} \nonumber
E_{\rm z,max} &=& eNk_p^2\exp\left(-\frac{k_p^2\sigma_z^2}{2}+\frac{k_p^2\sigma_r^2}{2}\right)\Gamma(0,k_p^2\sigma_r^2/2) \\
                        &=&eN Z(k_p,\sigma_z)R(k_p\sigma_r)
\end{eqnarray}
with
\begin{eqnarray}
 \label{e3}  Z(k_p,\sigma_z) &=&
    k_p^2\exp\left(-\frac{k_p^2\sigma_z^2}{2}\right),  \\
 \label{e4} R(k_p\sigma_r) &=&
    \exp\left(\frac{k_p^2\sigma_r^2}{2}\right)\Gamma(0,k_p^2\sigma_r^2/2),
\end{eqnarray}
where $\Gamma (\alpha, \beta) = \int_\beta^\infty t^{\alpha-1} e^{-t} dt$ is
the incomplete Gamma function, $e$ is the fundamental electric charge, $N$ is
the number of particles in the driving bunch, $\sigma_z$ is the rms length of
the driving bunch, $\sigma_r$ is the rms transverse size of the driving bunch
in one dimension (two transverse sizes are assumed equal), and $k_p=\omega_p/c$
is the plasma wavenumber determined by the plasma frequency $\omega_p$ and the
driver velocity that is close to the light velocity $c$. For a fixed bunch
length $\sigma_z$, the maximum of $Z$ is given by $k_p^2\sigma_z^2=2$. In our
considerations (modulation of a long bunch), $\sigma_z$ for a microbunch is
completely fixed by $k_p$ in such a way that that this condition is
approximately satisfied.

The function $R(k_p\sigma_r)$ peaks at small $k_p\sigma_r$  and can be
parametrized for $0.04<k_p\sigma_r<1.3$ to better than 10~\% accuracy as
 $$R(k_p\sigma_r)\approx \frac{7.2\exp(-k_p\sigma_r/5)}{1+5\,k_p\sigma_r} \;\; .$$
We see from these formulae that small transverse sizes of the proton bunch are
critical, as well as maximum plasma density.  We determine the optimum
conditions below.

\section{Modulated bunch}

Suppose we have $\sigma_{z,0}k_p\gg 1$, where $\sigma_{z,0}$ is the initial
length of the proton bunch as it enters the plasma.  Under the right
conditions, an instability will be quickly amplified resulting in microbunching
of the long bunch. The number of e-foldings of the initial instability
scales~\cite{ref:KP} as
\begin{eqnarray}
\label{eqn:foldings} N_e \approx 0.3\,\xi^{2/3}\tau^{1/3}
\end{eqnarray}
where $\xi$ is the distance along the bunch in the co-moving frame in units of
$k_p^{-1}$ and $\tau$ is the `time' the bunch has traveled in the plasma in
units $c/\omega_b$. Here
 $$
\omega_b=\sqrt{\frac{4\pi n_b e^2}{\gamma_bM}},
 $$
$M$ and $\gamma_b$ are the mass and relativistic factor of the driving bunch particles,
and $n_b$ is the typical bunch density that, in the central part of the
Gaussian bunch, is
\begin{equation}\label{e2}
  n_b = \frac{N}{(2\pi)^{3/2} \sigma_r^2 \sigma_z}.
\end{equation}
Note that this definition of $\tau$ differs from that in \cite{ref:KP} by the
factor of $\sqrt{2}$.

Table~\ref{tab:parameters}  gives the value of $c/\omega_b$ taking the peak
density for PS, SPS and LHC bunch options. Expression~(\ref{eqn:foldings}) is
applicable if $\tau$ is much greater than the e-folding time, which is not
strictly applicable to our case, but the regime characterized by
(\ref{eqn:foldings}) is the one we want to reach. As seen from the table, we
will need a very long plasma cell to reach large values of $\tau$.
\begin{table}[htdp]
\caption{PS, SPS and LHC parameter sets.  The different symbols are defined in
the text. SPS-LHC means the standard parameters of bunches in the SPS for
injection into the LHC.  SPS-Totem means the special parameters for bunches for
use by the Totem experiment.}
\begin{center}
\begin{tabular}{|c|c|c|c|c|}
\hline
 Parameter & PS & SPS-LHC & SPS-Totem & LHC \\
 \hline
 $W_P$ (GeV) & 24 & 450 & 450 & 7000  \\
 N$_P$ ($10^{10}$) & 13  & 11.5 & 3.0  & 11.5 \\
 $\sigma_{_P}$ (MeV) & 12  & 135 & 80 & 700  \\
 $\sigma_{z,0}$ (cm) & 20 & 12 & 8 & 7.6  \\
 $\sigma_r$ ($\mu$m) & 400 & 200 & 100 & 100  \\
 $c/\omega_b$ (m) & $2.3$ &  $4.0$ & $3.2$ & $6.3$  \\
 $\sigma_{\theta}$ (mrad) &  0.25 & 0.04 & 0.02 & 0.005  \\
 $L_{\theta}$ (m) & $1.6$ & $5$ & $5$ & $20$  \\
 $\epsilon$ (mm-mrad) & $0.1$ & $0.008$ & $0.002$ & $5\cdot 10^{-4}$  \\
 \hline
\end{tabular}
\end{center}
\label{tab:parameters}
\end{table}%

We would like to have a well developed modulation in the center of the bunch,
since this is where the bulk of the protons reside.  Assuming $\tau$ will be a
number of order $1$, we need to have large values of $\xi$ to make sure the
modulation is well developed.  This implies that the plasma wavelength should
be much shorter than the bunch length.  Indeed, we would want to maximize $k_p
\sigma_{z,0}$. However, we need the plasma skin depth to be at least as large
as the transverse size of the bunch, i.e.,
 $$\frac{c}{\omega_p}\geq \sigma_r$$
or
 $$k_p\sigma_r\leq1\;\; .$$
If this requirement is not imposed, filamentation of the bunch could likely
develop instead of the axisymmetric instability mode. For a fixed value of
$\sigma_r$, examination of the formulae in the previous section indicate that
the optimum fields under this requirement will be achieved for $k_p\sigma_r=1$,
and we will use this in the following calculations. The plasma frequency
$\omega_p$ is related to the plasma density $n_p$ as
 $$
\omega_p = \sqrt{\frac{4 \pi n_p e^2}{m}},
 $$
where $m$ is the electron mass. Fixing $k_p\sigma_r=1$ implies that the plasma
density is fixed once we fix the transverse size of the proton bunch, as are
the frequency $\omega_p$ and wavelength $\lambda_p$ of the plasma oscillations.
A useful relation is $$\lambda_p\approx 1~{\rm mm} \sqrt{\frac{10^{15} {\rm
cm}^{-3}}{n_p}}.$$ The values for the optimal density and plasma wavelength are
given in Table~\ref{tab:fields} for the different bunch parameters. The typical
plasma densities of interest will be around $10^{15}$~cm$^{-3}$ and the
resulting plasma wavelengths around $1$~mm.
\begin{table}[htdp]
\caption{Optimal values for the plasma density and resulting values of the parameters defined in the text.}
\begin{center}
\begin{tabular}{|c|c|c|c|c|}
\hline
 Parameter & PS & SPS-LHC & SPS-Totem & LHC \\
 \hline
 $n_p$ ($10^{15}\,\text{cm}^{-3}$) & $0.18$ & $0.7$ & $3.0$ & $3.0$  \\
 $\lambda_p$ (mm) & $2.5$ & $1.3$ & $0.6$ & $0.6$  \\
 $W_f$ (eV) & 180 & 280 & 100 & 410  \\
 $W_{tr}$ (eV) & 750 & 360 & 90 & 90  \\
 $eE_{\rm z,max}$ (GeV/m) & $0.1$ &  $0.35$ & $0.35$ & $1.4$  \\
 $eE_0$ (GeV/m) & $1.3$ & $2.5$ & $5.3$ & $5.3$  \\
 $\alpha$ & $0.08$ & $0.15$ & $0.07$ & $0.25$  \\
\hline
\end{tabular}
\end{center}
\label{tab:fields}
\end{table}%

Since the upper limit on $k_p$ is fixed from the bunch radius, it is seen that
maximizing $\xi$ implies maximizing the aspect ratio of the bunch
$\sigma_{z,0}/\sigma_r$.  In practice, there will be a limit to the number of
microbunches which will behave in a coherent manner due to phase slippage of
the instability relative to the bunch, plasma density variations, etc.  However,
the proton bunch should have a very large aspect ratio in order to reach the
stage of strong modulation.  This is the case for existing proton bunches,
including those listed in Table~\ref{tab:parameters}.
\begin{figure*}[htb]
 \bc\includegraphics[width=431bp]{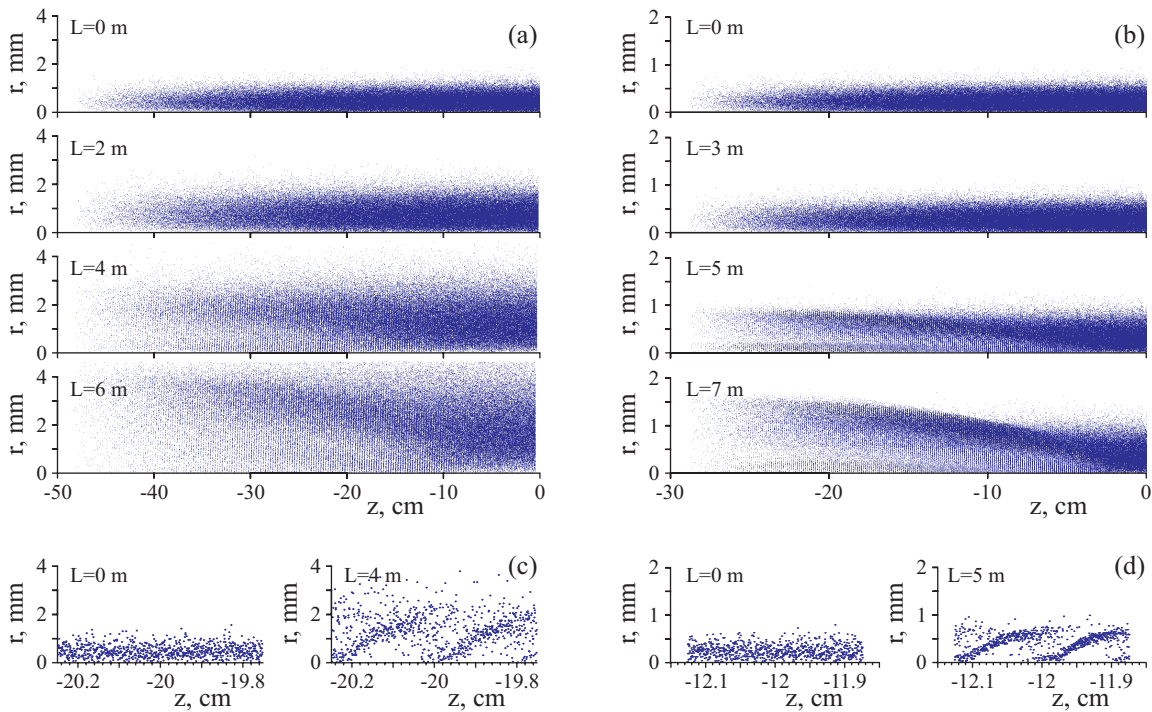} \ec
\caption{(Color online) Bunch portraits at different propagation lengths
showing the part of the bunch traveling in the plasma (a,b) and the zoomed area
located $\approx \sigma_z$ behind the bunch center (c,d) for the PS bunch
(left) and SPS-LHC bunch (right).}\label{f1-inst}
\end{figure*}

\section{Simulation Results for Instability Development}
The development of the modulation has been studied in simulations using both a
seeded and unseeded instability.  Only results for the seeded case are reported here.  Somewhat higher gradients were observed in simulations for the unseeded case, but we believe the results for the seeded instability are more reliable.

The proton bunch was simulated with a cosine profile given by
 $$
n_P = \frac{N_P}{2\sigma_r^2\sigma_z(2\pi)^{3/2}}
e^{-r^2/2\sigma_r^2}\left[1+\cos\left(\sqrt{\frac{\pi}{2}}\frac{z}{\sigma_z}\right)\right]
 $$
with the limit $|z|<\sigma_z\sqrt{2\pi}$.  For the seeded case, it was assumed
the plasma was created simultaneously with the passage of the proton bunch, and
started at the midpoint of the bunch (e.g., via a co-propagating laser
pulse~\cite{ref:CJ}). We used the 2D axisymmetric code LCODE with the fluid model
for plasma electrons \cite{PoP5-785,PRST-AB6-061301}. Since the beam-to-plasma
density ratio is less than 0.3\% for all considered variants, the plasma wave
does not break and can be safely described within the fluid approximation.
\begin{figure*}[htb]
 \bc\includegraphics[width=443bp]{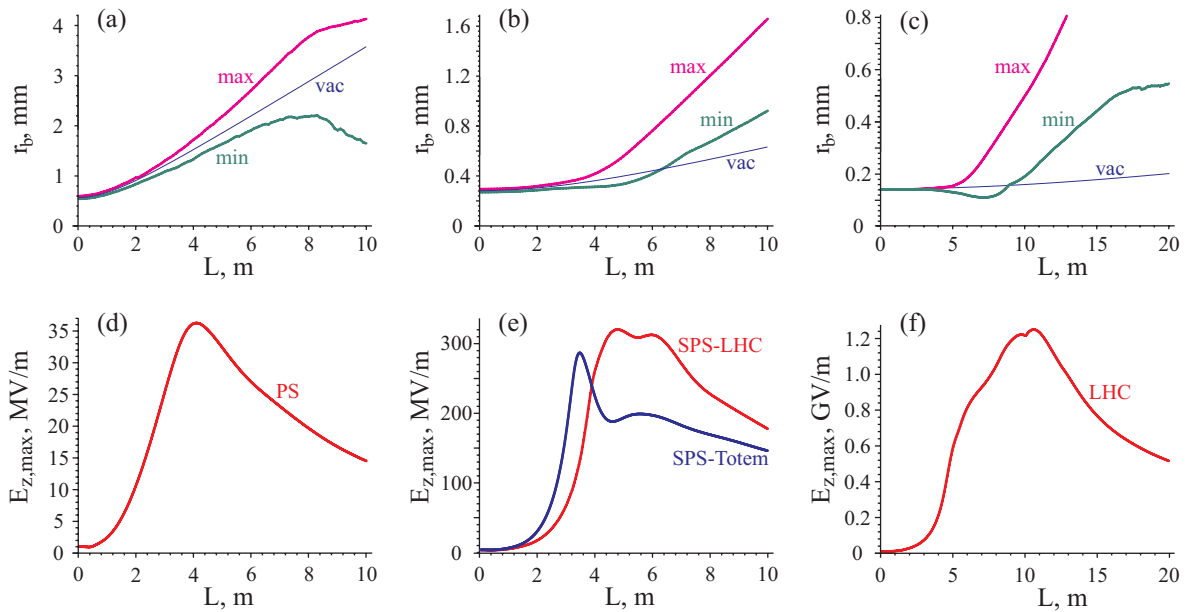} \ec
\caption{(Color online) Maximum and minimum rms radius of the proton bunch at
longitudinal distances $\approx \sigma_z$ behind the bunch center (a,b,c) and
the maximum wakefield amplitude (d,e,f) versus propagation distance: PS (a,d),
SPS-LHC (b,e), SPS-Totem (e), and LHC (c,f). The thin lines in (a,b,c)
correspond to free expansion of bunches in vacuum.}\label{f2-growth}
\end{figure*}

Figure \ref{f1-inst} shows the development of the seeded instability for PS and SPS-LHC bunches. Only the parts of the bunch propagating in the plasma are shown. In both cases we observe strong modulation of the bunch density and the typical trumpet-like structure of the disrupted bunch after several meters of propagation. There is, however, a qualitative difference between these two cases, which is more visible in Figure~\ref{f2-growth}. While the SPS-LHC bunch widens due to the instability, the divergence of the PS bunch is caused mainly by the angular spread of the beam particles. In the latter case, the instability develops against the background of this divergence and produces much lower fields as a result.

We consider bunch divergence more closely. The angular spread $\sigma_{\theta}$
of the bunch is related to its radius $\sigma_r$ via the emittance $\epsilon$:
 $$\sigma_{\theta}= \epsilon/\sigma_r \;\; .$$
The distance at which a particle shifts transversely by the distance of about
$\sigma_r$ is $L_{\theta} = \sigma_r/\sigma_{\theta} = \sigma_r^2/\epsilon$;
that is, the betatron function of accelerator physics. Values of $L_{\theta}$
are listed in Table~\ref{tab:parameters}. The ratio $L_{\theta} \omega_b/c$ is
greater than 1 for the PS bunch, and smaller than 1 for the SPS bunch, which
explains the qualitative difference between the two cases.

There is another effect that could be important for the development of the
instability; transverse focusing by plasma fields. To estimate the effect, we
compare the energy associated with the transverse motion of protons and the
depth of the potential well that focuses the bunch. For long and narrow bunches
($\sigma_r \lesssim c/\omega_p$, $\sigma_z \gg c/\omega_p$) the focusing force
is mainly determined by the uncompensated current of the beam. The focusing
strength is $F_r/r \sim 2 \pi n_b e^2$, from which we estimate, for a bunch of
radius $\sigma_r \sim c/\omega_p$, the depth of the potential well to be
\begin{equation}\label{e8}
    W_f \sim \frac{F_r \sigma_r}{2} \sim \frac{m c^2 n_b}{4 n_p} \;\; .
\end{equation}
The energy of transverse motion $W_{tr}$ is determined by the transverse
momentum $p_{b\perp}$ of bunch particles:
\begin{equation}\label{e9}
 W_{tr} = \frac{p_{b\perp}^2}{2 \gamma_b M}
\approx \frac{\sigma_{\theta}^2 W_b^2}{2 \gamma_b M c^2} = W_b
\frac{\sigma_{\theta}^2}{2} = W_b \frac{\epsilon^2}{2 \sigma_r^2},
\end{equation}
where $W_b$ is the incident proton energy. The numerical values are shown in Table~\ref{tab:fields}. For the PS bunch, plasma focusing is negligible. For both variants of the SPS bunch, the effect of focusing is visible (some parts of the bunches diverge more slowly than in vacuum), but not strong. For the LHC bunch, the plasma focusing is important and favorable for the instability. The bunch is pinched by the plasma wave at some cross sections, which results in stronger wakefields.  This pinching is visible after $5-10$~m of propagation in the plasma.  At longer distances, the microbunch structure is perturbed by the instability as discussed below.

\section{Fields from maximum modulation}

We now suppose we can reach the state where we have maximum modulation of the
bunch, and investigate the strength of the electric fields.  The bunch has been divided into
microbunches which each contains $\approx 1/2$ of the initial number of protons
within one plasma wavelength (the other protons are pushed out by the radial
fields).  We can use the expression in the first section to estimate the
electric field from one microbunch:
\begin{equation}
E_{\mu, \rm z,max}  = eN_{\mu} Z(k_p,\sigma_z)R(k_p\sigma_r)
\end{equation}
where $N_{\mu}$ is the number of protons in the microbunch and $\sigma_z\approx
\sqrt{2}k_p^{-1}$ is the rms length of the protons in the microbunch.  If we
assume for seeded instabilities that all microbunches behind the center of the
proton bunch add coherently to the produced electric field, then we have
\begin{equation}
E_{\rm z,max} \approx \frac{e N}{4} Z(k_p,\sigma_z)R(k_p\sigma_r) \; .
\end{equation}
We now calculate the maximum electric field by taking $k_p\sigma_r=1$,
substituting $\sigma_z\approx \sqrt{2} k_p^{-1} = \sqrt{2}\sigma_r$, and using
(\ref{e3}), (\ref{e4}). This yields
\begin{equation} \label{e11}
E_{\rm z,max}   \approx  0.085 \, \frac{N e}{\sigma_r^2} \approx 0.12 ({\rm
GV/m}) \cdot \left(\frac{N}{10^{10}}\right) \left(\frac{100~\mu{\rm
m}}{\sigma_r}\right)^2.
\end{equation}
The maximum field from this expression is given in Table~\ref{tab:fields}. The
fields can be compared to the wave-breaking field
 $$eE_0 = k_p mc^2 $$
to determine the dimensionless field amplitude
\begin{equation}
\alpha=\frac{E_{\rm z,max}}{E_0} \approx 0.024 \left(\frac{N}{10^{10}}\right)
\left(\frac{100~\mu{\rm m}}{\sigma_r}\right)\;\; .
\end{equation}
As is seen in the table, the values of $\alpha$ range from 7~\% for the
SPS-Totem to 25~\% for the LHC. Estimate (\ref{e11}) agrees well with
simulation results (Fig.\,\ref{f2-growth}) for all variants except PS. For the
PS bunch, the simulated field is lower because of the angular divergence of the
bunch which is not taken into account in (\ref{e11}).  As is clear from the
formulae in this section, the critical element determining the strength of the
wakefields in the modulated bunch case is the transverse density of particles
in the drive bunch,  $N/\sigma_r^2$,  so
that a reduction of the transverse emittance or an increase in the bunch
intensity will have significant effect on the achievable accelerating fields.

\section{Evolution of the microbunches}

\begin{figure}[htb]
 \bc\includegraphics[width=223bp]{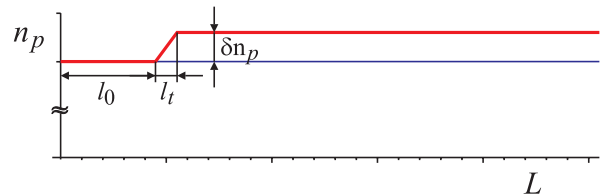} \ec
\caption{(Color online) Plasma density profile used in the simulation for the
creation of a long-lived train of microbunches.}\label{f3-step}
\end{figure}

There are several effects limiting the distance over which the strong fields act to increase the
energy of accelerated particles. If the drive bunch propagates in a uniform
density plasma, then the instability destroys the microbunches soon after the maximum
field is reached. This effect is clearly seen in Fig.\,\ref{f2-growth} for all
four variants. The reason lies in the slow motion of the defocusing field regions with
respect to the bunch. This motion continues after the bunch formation is completed and
quickly destroys the microbunches \cite{TTScontrol}. It is possible to avoid
the destruction of the microbunch structure by a proper step up in the plasma density
(Fig.\,\ref{f3-step}) which modifies the instability growth in such a way that
the field motion relative to the bunches stops at the optimal moment
\cite{TTScontrol}.  The density step cannot stop the development of the instability immediately. It quickly changes the excited field, but the radially moving protons still have inertia. Thus the density step modifies the instability in such a way that the bunch evolves toward the desired state and the evolution finishes at this state.
The values for the parameters $\delta n_p$, $l_0$, and $l_t$ maximizing the
established wakefield amplitude are listed in Table~\ref{t3}.
\begin{table}[htdp]
\caption{Parameters describing the bunch interaction with the plasma in the case of the stepped-up plasma density.}
\begin{center}
\begin{tabular}{|c|c|c|c|c|}
\hline
 Parameter & PS & SPS-LHC & SPS-Totem & LHC \\
 \hline
 $\delta n_p$ (\%)     & $2.2$ & 2.4 & $2$ & $1.6$ \\
 $l_0$ (m)             & $1$   & 2.5 & $1$ & $3$   \\
 $l_t$ (m)             & $3$   & 1.5 & $3$ & $1.5$ \\
 $N_{mb}$              & 80    & 95  & 130 & 125   \\
 $L_\text{diffr}$ (m)  & 130   & 475 & 655 & 2500  \\
 $E_{\rm z,max}$ (MV/m)& 13    & 100 & 100 & 900   \\
 $L_\text{depl}$ (km)  & 1.8   & 4.5 & 4.5 & 7.8   \\
 $W'$                  &0.00012&0.009&0.004& 0.6   \\
 $L_\text{dec}$ (m)    & 32    & 670 & 470 & 9500  \\
 $W_\text{dec}$ (GeV)  & 0.4   &  60 & 40  & 7700  \\
 $W_E$ (GeV)           & 0.25  &  10 & 10  & 10    \\
 $L_\text{deph}$ (m)   & 1.4   & 250 & 120 & 30000 \\
 $W_\text{deph}$ (GeV) & 0.012 &  16 &  8  & 17000 \\
\hline
\end{tabular}
\end{center}
\label{t3}
\end{table}%
\begin{figure*}[htb]
 \bc\includegraphics[width=444bp]{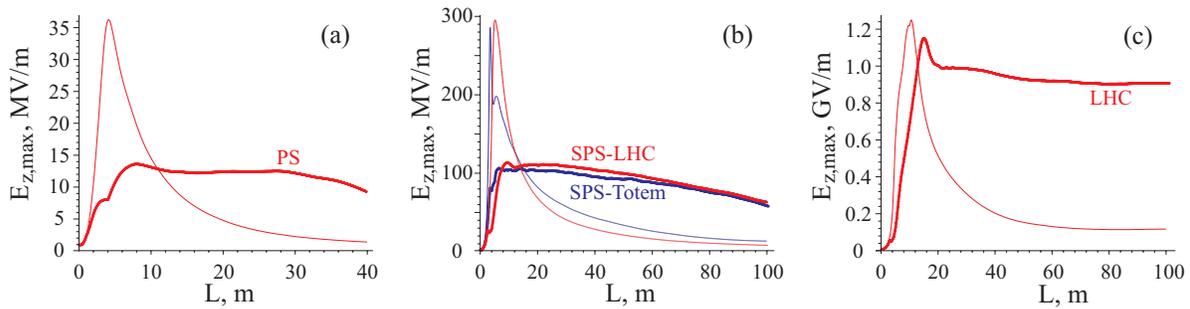} \ec
\caption{(Color online) The maximum wakefield amplitude behind the bunch center
versus the propagation distance for the stepped-up (thick lines) and uniform
(thin lines) plasmas.}\label{f4-midterm}
\end{figure*}

The field evolution for the stepped plasma profile is shown in
Fig.\,\ref{f4-midterm} in comparison with the uniform plasma case. The
wakefield is preserved for a long distance, but however has a lower
amplitude than in the uniform case. The latter follows from the fact that
proton microbunches can create a wakefield over a long distance only if they are located
in both the decelerating and the focusing phase of the wakefield, which is only one
quarter of the wakefield period. In the uniform plasma, the maximum of the field
occurs when the microbunches fill the whole decelerating phase of the
wave, which is nearly one half of the period. Thus, the field amplitude for the
long-lived microbunch train should be roughly half that of the peak field in the
uniform plasma. Simulations confirm this statement for all variants except LHC
bunch. For the latter, the field of the long-lived train is substantially increased
due to the plasma focusing of the microbunches with respect to their initial size.

\begin{figure*}[htb]
 \bc\includegraphics[width=443bp]{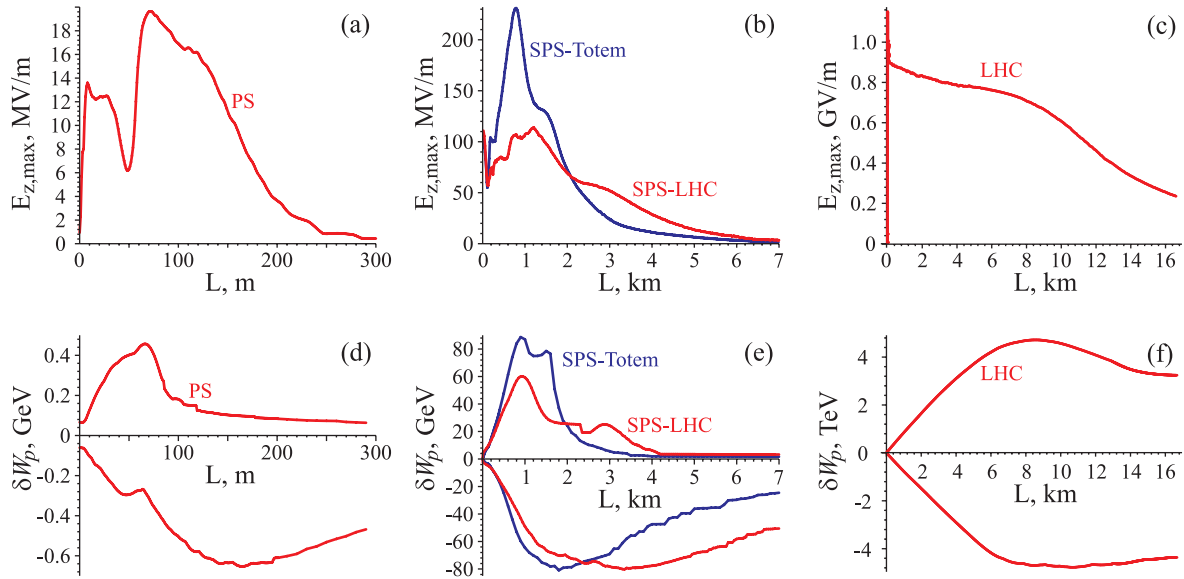} \ec
\caption{(Color online) The dependence of the wakefield amplitude (top) and
maximum proton energy gain or loss (bottom)  over long distances for the
stepped-up plasma case.} \label{f5-field}
\end{figure*}
While most of microbunches in the long-lived train are well focused by the
plasma wave, the head of the first microbunch is not and doubles in size
transversely after a distance $\sim L_{\theta}$. Once the first microbunch has
diffracted away, the next will start to diffract, and this process is expected
to continue until all microbunches are lost. The lower estimate for the scale
over which, in the absence of external focusing, the whole bunch diffracts
away, is
 $$L_\text{diffr} = N_{mb} L_{\theta},$$
where $N_{mb} \sim \sigma_{z,0}/\lambda_p$ is the number of microbunches. Values
of $N_{mb}$ and $L_\text{diffr}$ are given in Table~\ref{t3}. However,
simulations reveal that the field excitation is preserved over a much longer length
[Fig.\,\ref{f5-field}(b,c)] suggesting that only a small fraction of a microbunch becomes unavailable for the
wakefield excitation at each path segment of the length $L_{\theta}$.

\begin{figure*}[htb]
 \bc\includegraphics[width=390bp]{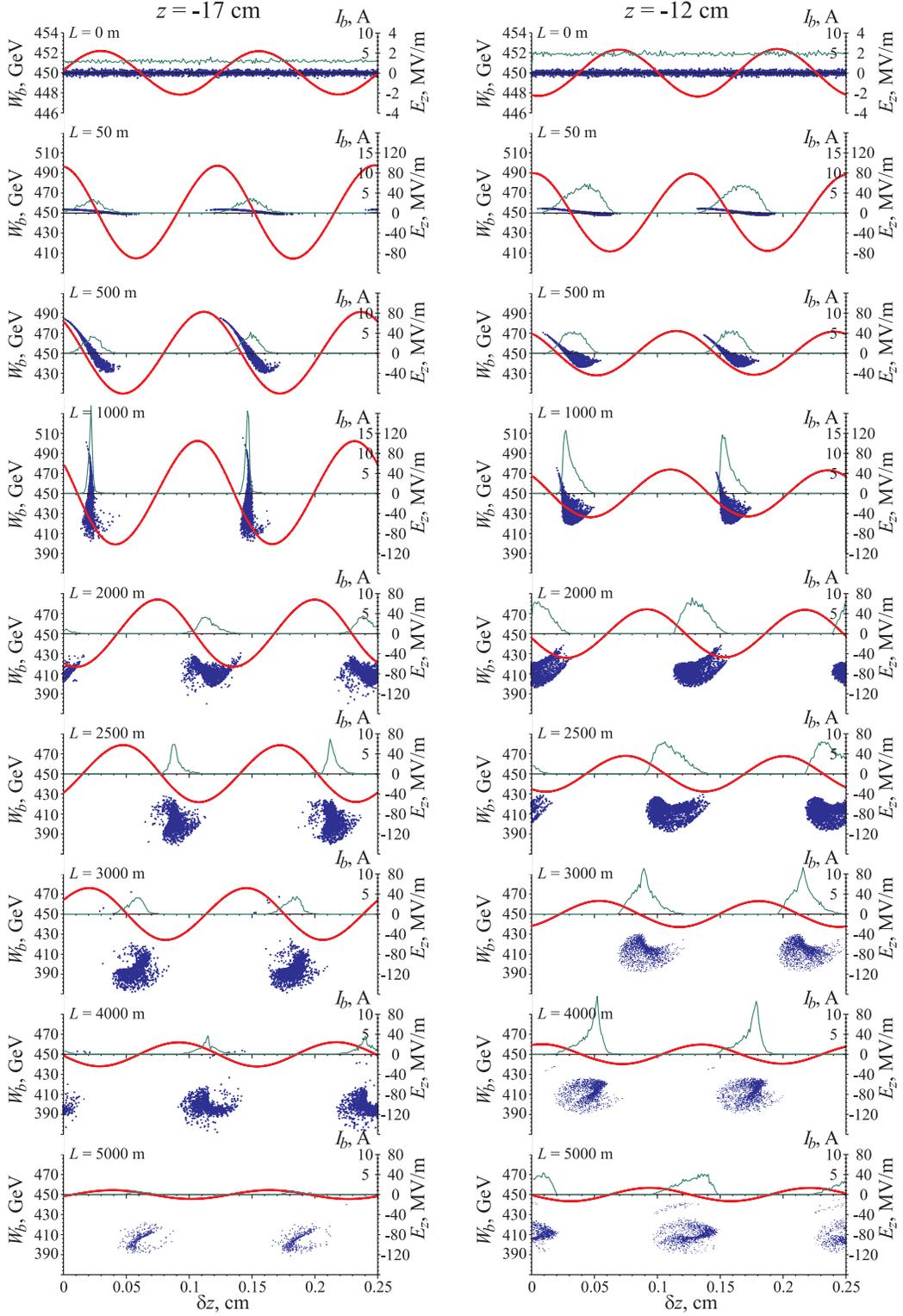} \ec
\caption{(Color online) Evolution of several SPS-LHC microbunches: energy $W_b$
of bunch particles (dots), on-axis electric field $E_z$  (red line), and bunch
current $I_b$ through the circle of radius $c/\omega_p$ (green line) vs
longitudinal coordinate $\delta z$ at various distances traveled in the
stepped-up plasma. The coordinates $\delta z$ are measured from points moving
with the speed of a $450$~GeV proton and initially located $17$~cm (left
column) or $12$~cm (right column) behind the bunch center.}\label{f6-phase}
\end{figure*}

The second possible limitation on the length of the strong wakefield excitation
comes from the energy change of protons in the bunch. The simplest consideration of this effect,
estimated by dividing the proton initial energy by the decelerating force,
\begin{equation}\label{e13}
  L_\text{depl} = \frac{W_P}{e E_{\rm z,max}},
\end{equation}
typically overestimates the length. The distances $L_\text{depl}$ based on
simulated values of $E_\text{z,max}$ (Table~\ref{t3}) are much longer than the
lengths over which the fields decrease in Fig.\,\ref{f5-field}(a,b). A more
careful investigation shows that the difference in energy depletion among the
protons within a microbunch must be taken into account. For short proton
bunches \cite{PRST-AB13-041301}, the energy depletion results in an elongation
of the bunch and thereby gradual decrease of the excited wakefield. For trains
of microbunches, the mechanism is different, as illustrated by
Fig.~\ref{f6-phase}. The microbunches are initially located in such a way with
respect to the wakefield that the microbunch head experiences the strongest
deceleration. The induced energy chirp along the bunch results in bunch
compression and a net shift of the bunch to the point of zero longitudinal
field. When this occurs, the bunch stops changing its average energy and
contributing to field excitation. When most of the bunches are in this
situation,  the wakefield drops down, and the bunches get lost transversely
since the weaker wakefield cannot keep the protons focused any more.

To estimate the energy change of a microbunch with distance, we assume a constant wave amplitude and consider the process in the
reference frame moving with the phase speed of the wave $V \approx c$. The wakefield
$E_z'$ and the on-axis potential $\Phi'$ in this frame are
\begin{equation}\label{e14}
  E_z'= -E_{\rm z,m} \sin k_p' z', \qquad
  \Phi' = -\frac{e E_{\rm z,m}}{k_p'} \cos k_p' z',
\end{equation}
where $E_{\rm z,m}$ is the on-axis field amplitude in the laboratory frame,
$k_p'=\omega_p/(\Gamma V)$ is the wavenumber in the moving frame, \ $\Gamma =
(1-V^2/c^2)^{-1/2}$, and the coordinate $z'$ is measured from a zero field
point. In the potential well (\ref{e14}), a proton starting from the point of
the strongest field ($k_p' z'=\pi/2$) gains a maximum kinetic energy (in units
of the rest energy) of
\begin{equation}\label{e15}
    W' = \frac{e E_{\rm z,m}}{k_p' M c^2}.
\end{equation}
Values of  $W'$ for $E_{\rm z,m}=E_{\rm z,max}$ are listed
in Table~\ref{t3}. For the first three variants, $W' \ll 1$, and the
motion of bunch particles in the moving frame is not relativistic. The frequency
of particle oscillations near the bottom of the potential well is thus
independent on the amplitude and equals
\begin{equation}\label{e16}
 \omega_{z'} = \sqrt{\frac{e E_{\rm z,m} k_p'}{M}}.
\end{equation}
Simultaneous events separated in the laboratory frame by the distance
$c/\omega_p$ (i.e., by the microbunch length) differ in the moving frame by the time interval $(k_p'
c)^{-1}=\sqrt{W'} / \omega_{z'}$. For $W' \ll 1$, we can
neglect this time difference and consider the bunch particles to start their motion
simultaneously in the moving frame. Consequently, after the time $\pi/(2 \omega_{z'})$
in the moving frame and the distance
\begin{equation}\label{e17}
 L_\text{dec} = \frac{\pi c \Gamma}{2 \omega_{z'}}
 \approx \frac{\pi c \Gamma^{3/2}}{2 \omega_p} \sqrt{\frac{M E_0}{m E_{\rm z,m}}}
\end{equation}
in the laboratory frame, a considerable fraction of the protons collect
at the bottom of the potential well (Fig.~\ref{f6-phase}, $L=1000$\,m). This is
the typical length scale over which the microbunch gives up the energy to the
wave, if the bunch was initially located in the decelerating phase of the
wakefield. To obtain the energy scale, we Lorentz transform the energy
(\ref{e15}) and the corresponding momentum to yield the energy loss of protons
\begin{equation}\label{e18}
  W_\text{dec} = W_P \sqrt{2 W'}.
\end{equation}
For the LHC bunch ($W'=0.6$), the above formulae are marginally correct, but still
can be used as order-of-magnitude estimates. Values of $L_\text{dec}$ and
$W_\text{dec}$ are given in Table~\ref{t3}. For the LHC case, $W_\text{dec}>W_P$,
implying that the longitudinal shift of depleted particles is not a limiting factor.

\begin{figure*}[htb]
 \bc\includegraphics[width=439bp]{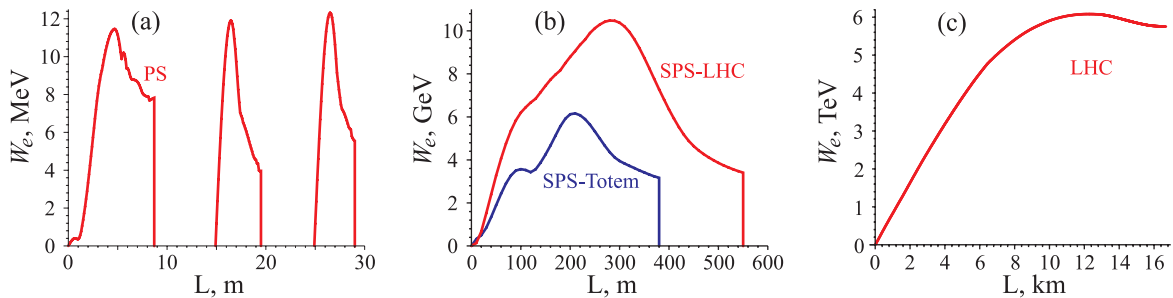} \ec
\caption{(Color online) The maximum energy gain or electrons versus propagation
distance for the stepped-up plasma case.}\label{f7-electrons}
\end{figure*}
The real picture of wave-microbunch interactions is more complicated than that
described in our simplified model, as can be seen in Fig.~\ref{f6-phase}. Since
the wave is created by the microbunches themselves, the cross-section of zero
$E_z$ moves backward together with upstream microbunches, and compressed
microbunches are formed in a decelerating field rather than at the point of
zero acceleration. As a consequence, the distance of strong field excitation is
longer than $L_\text{dec}$ (Fig.~\ref{f5-field}, top), the energy loss of
protons is greater than $W_\text{dec}$ (Fig.~\ref{f5-field}, bottom), and
protons do not make a full circle on the longitudinal phase plane taking the
energy back from the wave, but instead stick to the zero-field point and stop
energy exchange with the wave (Fig.~\ref{f6-phase}).

\section{Electron energy gain}

If test electrons of energy $W_E \gg m c^2$ are injected into the
wakefield, their energy gain (Fig.\,\ref{f7-electrons}) is much lower than the
energy gain of the most energetic protons for all variants except the LHC
[Fig.\,\ref{f5-field}(d-f)]. The reason is the dephasing between electrons and the
wakefield. The typical distance at which a fast electron outruns a proton with the
relativistic factor $\Gamma$ by the half of the wakefield period (and crosses
the whole acceleration phase of the wave) is
\begin{equation}\label{e19}
    L_\text{deph} = 2 \pi \Gamma^2 c/\omega_p.
\end{equation}
To obtain the typical energy gain we multiply this distance by the average
accelerating field:
\begin{equation}\label{e20}
   W_\text{deph} = \frac{2}{\pi} e E_\text{z,max} L_\text{deph} \; .
\end{equation}
The resulting values are given in Table~\ref{t3}. The parameters obtained in
simulations are close to (\ref{e19})--(\ref{e20}). For the PS case, electrons
were injected into the wakefield several times at various stages of bunching
[Fig.\,\ref{f7-electrons}(a)], but for all groups the maximum electron energy
is the same.

To obtain higher energy electrons from the PS or SPS energy proton bunches, it would be necessary to achieve larger gradients. This would require larger drive bunch density, either through lower emittances or larger numbers of drive particles.

\section{Limitations due to interactions}
As discussed in \cite{NatPhys9-363}, proton interactions in the plasma are not a limiting factor.  For typical values of $n_p$ in the range $10^{14}-10^{16}$~cm$^{-3}$, the mean-free-path for inelastic reactions of high energy protons with the plasma are orders of magnitude larger than the simulated plasma cell lengths.  The transverse growth rate of the proton bunch due to multiple scattering is also very small compared to other effects discussed here.  For electrons, the radiation lengths for the plasma densities of interest here are also very long compared to the lengths of the plasma cells under consideration, so that particle interactions in the plasma are not expected to be a limiting factor.

\section{Summary}

We have investigated the wakefield amplitudes which could be achieved via the
modulations of a long proton bunch.  The key result in the parametric
investigation is that, in the limit of long bunches compared to the plasma
wavelength, the strength of the accelerating fields is directly proportional to
the transverse particle density in the drive bunch, $N/\sigma_r^2$.  This finding puts a
premium on increasing this quantity for beams designed specifically for beam
driven wakefield acceleration via the modulation technique.

The scaling laws were tested and verified in detailed simulations using
parameter sets for existing or planned  proton bunches at CERN; i.e., PS, SPS
and LHC bunch parameters.  For existing bunch parameters, it was found that
electric fields of  about $35, 300, 1200$~MeV/m should be achievable for the
PS, SPS and LHC bunches, respectively.  These gradients cannot be maintained
over arbitrarily long distances because the modulation process will eventually
destroy also the microbunch structure. To overcome this deleterious effect, it
has been shown that a density step in the plasma can freeze the microbunch
structure, albeit at a penalty in the maximum achievable gradients.

A detailed investigation of the evolution of the microbunches was undertaken
for the setpped-up plasma density scenario.  It was found that significant
accelerating gradients could be maintained over long distances, although the
resulting energy gains for electrons were moderate due to dephasing effects for
the PS and SPS cases.  For the LHC, the protons are relativistic enough that
dephasing is not a significant issue, and in this case test electrons were
accelerated beyond $6$~TeV with gradients approaching $1$~GeV/m.  It is
anticipated that the achieved energy gains could be substantially more
favorable if the transverse density of protons could be improved for all
scenarios considered.

\section{Acknowledgements}

The authors are grateful to C. Huang, A. Pukhov, O. Reimann, J. Viera and G.
Xia for stimulating discussions. Different parts of this work are supported by
RFBR Grants 09-02-00594, 11-01-00249, 11-02-00563, 11-02-91330, and by the
Russian Ministry of Education grants 2.1.1/3983 and 14.740.11.0053.

\end{document}